\documentclass[preprint,12pt]{elsarticle}

\usepackage{amssymb}
\usepackage{amsmath}

\journal{Physics Letters A}

\begin{document}

\begin{frontmatter}



\title{Coherence resonance in an excitable potential well}


\author{Tatiana R. Bogatenko}

\address{Department of Physics, Saratov State University, Astakhanskaya str. 83, 410012 Saratov, Russia}

\author{Vladimir V. Semenov}

\address{Department of Physics, Saratov State University, Astakhanskaya str. 83, 410012 Saratov, Russia}

\begin{abstract}
The excitable behaviour is considered as motion of a particle in a potential field in the presence of dissipation. The dynamics of the oscillator proposed in the present paper corresponds to the excitable behaviour in a potential well under condition of positive dissipation. Type-II excitability of the offered system results from intrinsic peculiarities of the potential well, whose shape depends on a system state. Concept of an excitable potential well is introduced. The effect of coherence resonance and self-oscillation excitation in a state-dependent potential well under condition of positive dissipation are explored in numerical experiments.
\end{abstract}


\end{frontmatter}


\section{Introduction}
The phenomenon of coherence resonance was originally discovered for excitable systems \cite{gang1993,pikovsky1997,lindner1999,lindner2004,deville2005,muratov2005} and then for non-excitable  ones \cite{ushakov2005,zakharova2010,zakharova2013,geffert2014}. This effect implies that noise-induced oscillations become more regular for an optimal value of the noise intensity. Coherence resonance is a frequent occurrence in neurodynamics \cite{pikovsky1997,lee1998,lindner2004,tateno2004} as well as in microwave \cite{dmitriev2011} and semiconductor \cite{hizanidis2006,huang2014} electronics, optics \cite{giacomelli2000,avila2004,otto2014,arteaga2007,arecchi2009}, thermoacoustics \cite{kabiraj2015}, plasma physics \cite{shaw2015}, and chemistry \cite{miyakawa2002,beato2008,simakov2013}. 

The noise-induced dynamics of excitable systems in the regime of coherence resonance is similar to the self-oscillatory behaviour. The similarity is complemented by the fact that noise-induced oscillations corresponding to the coherence resonance can be synchronized mutually or by external forcing \cite{han1999, ciszak2003,ciszak2004}. Moreover, the synchronization of the noise-induced oscillations occurs in a similar way as for a deterministic quasiperiodic system \cite{astakhov2011}. 

\begin{figure}
\begin{center}
\includegraphics[width=0.8\columnwidth]{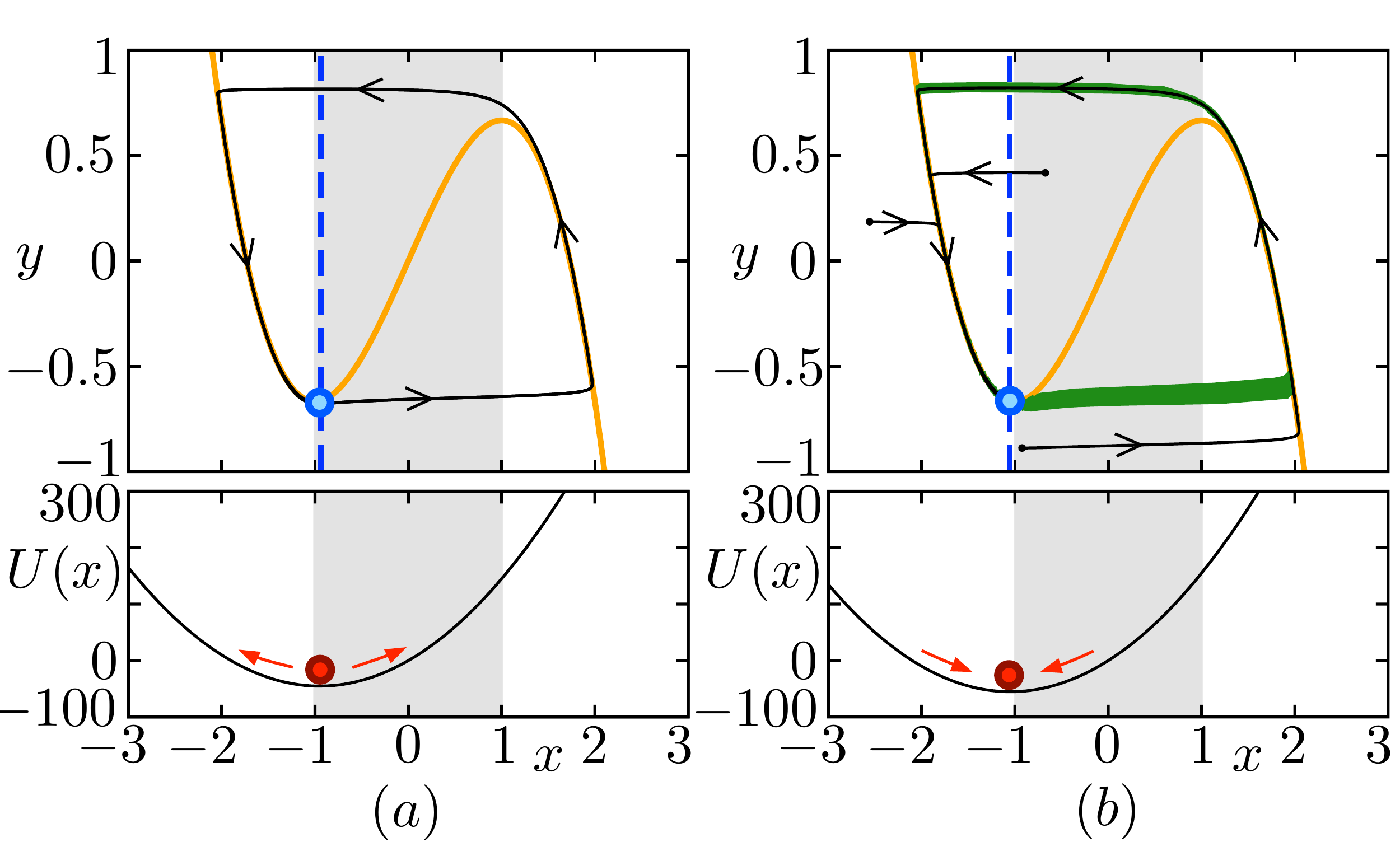}
\end{center}
\caption{Self-oscillatory (a) and excitable (b) regimes of the system (\ref{fhn}) in numerical simulation: phase space structure (upper panels) and corresponding potential function $U(x)$ (lower panels). The equilibrium point is shown by the blue circle; the blue dashed line indicates the nullcline $\dot{y}=0$; the orange solid line shows the nullcline ${\dot{x}=0}$. Phase trajectories of the deterministic system (in case $\xi(t)\equiv0$) are shown by black arrowed lines. Intrawell oscillations of a particle (the red circle on the lower panels) in the absence of noise are schematically shown by the red arrowed lines. The area corresponding to negative values of the dissipation function $\gamma(x)$ is  colored in gray. The green trajectory on the panel (b) corresponds to the system driven by white Gaussian noise ($\xi(t)=\sqrt{2D}n(t)$, $<n(t)> = 0$, $<n(t)n(t+\tau)> =\delta(\tau)$, $D$ is the noise intensity, $D=10^{-3}$). Parameters are: $\varepsilon=0.01$, $a=0.95$ (self-oscillatory regime), $a=1.05$ (excitable regime).}
\label{fig1}                                                                                                   
\end{figure}

Similarity of nature of self-oscillation excitation and excitability is also seen in the context of interpretation of the dynamics as motion of a particle in a potential field in the presence of dissipation. In such a case, a mathematical model of the system under study is presented in the following oscillatory form:
\begin{equation}
\label{general-eq}
\dfrac{d^{2}y}{dt^{2}}+\gamma\dfrac{dy}{dt}+\dfrac{dU}{dy}=0,
\end{equation}
where the factor $\gamma$ characterizes the dissipation, $U$ is a function denoting the potential field. Mathematically, a paradigmatic model for type-II excitability is the FitzHugh-Nagumo system \cite{fitzhugh1961,nagumo1962}:
\begin{equation}
\label{fhn}
\left\lbrace
\begin{array}{l}
\varepsilon \dfrac{dx}{dt}=x-\dfrac{x^3}{3}-y,\\
\\
\dfrac{dy}{dt}=x+a+\xi(t),
\end{array}
\right.
\end{equation}
where the parameter $\varepsilon$ sets separation of slow and fast motions, the parameter $a$ determines the oscillatory dynamics, $\xi(t)$ is a source of noise. In the oscillatory form (\ref{general-eq}) the model (\ref{fhn}) becomes:
\begin{equation}
\label{fhn-osc}
\varepsilon\dfrac{d^{2}x}{dt^{2}}+\left( x^2-1 \right)\dfrac{dx}{dt}+x+a=-\xi(t),
\end{equation}
and describes stochastic motion of a point mass in the potential $U(x)=\frac{x^2}{2\varepsilon}+\frac{a}{\varepsilon}x$ in the presence of dissipation defined by the function $\gamma=\gamma(x)=\frac{x^2-1}{\varepsilon}$. In case $|a|<1$ the dissipation function $\gamma(x)$  possesses negative values in the vicinity of a local minimum of a potential well. It denotes energy pumping, which leads to instability of an equilibrium point, and self-sustained oscillation excitation occurs [Fig.~\ref{fig1}~(a)]. Energy balance between dissipation and pumping during the period of the self-oscillations is organized after short transient time. 
The same self-oscillation excitation principle works in the Van der Pol \cite{vanderpol1920} and Rayleigh \cite{rayleigh1883} systems and in other examples of self-oscillators realizing the Andronov-Hopf bifurcation as well as in excitable systems with type-I excitability, where transition to the self-oscillatory regime is associated with the saddle-node bifurcation (see for example the two-dimensional modification of the Hindmarsh-Rose neuron model \cite{hindmarsh1984}). If changing of the parameter $a$ in the system (\ref{fhn}) shifts the equilibrium point out of the negative dissipation area [gray areas in Fig.~\ref{fig1}], then the steady state becomes stable and the system does not exhibit the self-oscillatory behaviour [Fig.~\ref{fig1}~(b)]. However, in the presence of noise the force $\xi(t)$ randomly kicks the phase point out of the vicinity of the stable equilibria towards the region of the negative friction $\gamma(x)$. Phase point drift is amplified in the areas of negative dissipation. If energy pumping is sufficient, the phase point trajectory forms a loop [the green trajectory in Fig.~\ref{fig1}~(b)]. Thus, both self-oscillation excitation and excitability are associated with the presence of negative dissipation\footnote{It is important to note that the definition ''positive'' or "negative dissipation" determined by the sign of the term $\gamma$ in the oscillatory forms (\ref{general-eq}) and (\ref{fhn-osc}) is correct only in terms of description of the dynamics as motion of a particle in a potential field. Generally, dissipativity of dynamical systems is determined by the divergence of the phase space flow.} in the context of motion of a particle in a potential field. 

The presented above interpretation of the self-oscillatory dynamics in terms of motion of a particle in a potential field involves the presence of a potential with a fixed profile and state-dependent dissipation, which possesses negative and positive values and is responsible for the existence of the self-sustained oscillations. Another configuration of self-oscillatory motion in the potential field is proposed in the paper \cite{semenov2018}. It implies positive dissipation factor and a state-dependent potential, which is responsible for self-oscillation excitation. In that case both steady state instability and amplitude limitation are dictated by the potential function. This configuration corresponds to mutual interaction of the point mass particle and the potential field. The potential determines the particle's dynamics, but it changes depending on the velocity of the mass point. The publication \cite{semenov2018} is focused on the dynamics of deterministic systems and does not involve study of the stochastic behaviour. The next step, which allows to expand a manifold of possible phenomena in the state-dependent potential well  under condition of positive dissipation, is consideration of noise-induced effects. In particular, the effect of coherence resonance can be explored.

In the current paper we propose an excitable oscillator being similar to the model offered in the paper \cite{semenov2018}. The dynamics of the system under study can be interpreted as motion of a particle in a state-dependent potential well in the presence of positive dissipation. The explored system exhibits the effect of coherence resonance associated with the type-II excitability. In contrast to the system (\ref{fhn}) and other systems the excitable dynamics of the system under study is fully defined by the nonlinearity of a potential function and does not result from properties of a dissipation function. Therefore, the observed behaviour can be interpreted as motion of a particle in an excitable potential well. In the present paper we demonstrate that the effect of coherence resonance  can be achieved in the state-dependent potential well under condition of positive dissipation. By this way we complement the results of the paper \cite{semenov2018}.

The present paper is organized as follows. In the section \ref{mm} the studied system is described in details. Then the effect of coherence resonance is shown in the noise-driven system (section \ref{cr}). Conclusions are presented in the section \ref{conclusions}.
\section{Model and Methods}
\label{mm}

\begin{figure}[t]
\begin{center}
\includegraphics[width=0.9\columnwidth]{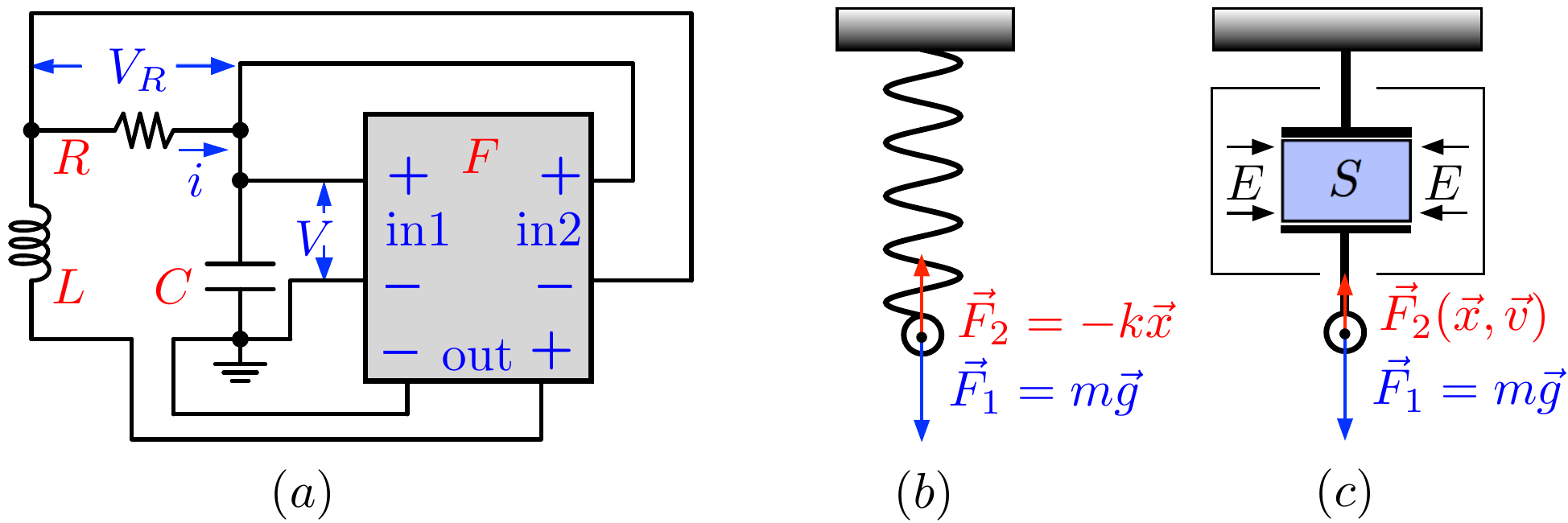}
\end{center}
\caption{(a) Schematic circuit diagram of the system under study (Eq.~(\ref{physical})). (b) Model of linear spring pendulum. (c) Modified spring-like pendulum based on elastic medium block.}
\label{fig2}                             
\end{figure}

Fig.~\ref{fig2} (a) shows a schematic circuit diagram, which mathematical model is derived below. It is a series-oscillatory LCR-circuit driven by a nonlinear feedback. The nonlinear feedback is realized by the nonlinear converter $F$. The converter $F$ has two inputs $V$ (the voltage across the capacitor $C$) and $V_{R}$ (the voltage across the resistor $R$) with zero input current. In that case the same current $i$ passes through the inductor $L$, the resistor $R$, the capacitor $C$ and then $V_{R}=iR$. The converter $F$ has the output voltage as being: $V_{out}=\frac{V_g}{(V-kV_{R})^2+1}-V_{a}+\xi(t)=\frac{V_g}{(V-m i)^2+1}-V_{a}+\xi(t)$, where $m=kR$, $V_{a}=$const, $V_{g}=$const, $\xi(t)$ is the  stochastic term determined by a broadband noise source included into the converter $F$. By using the Kirchhoff's voltage law the following differential equations for the voltage $V$ across the capacitor $C$ can be derived:
\begin{equation}
\label{physical}
\begin{array}{l}
CL\dfrac{d^2V}{dt'^2}+RC\dfrac{dV}{dt'}+V\\
\\
=\dfrac{V_g}{\left(V-m \frac{dV}{dt'}\right)^2+1}-V_{a}+\xi(t).
\end{array}
\end{equation}

In the dimensionless form with variable $x=V / v_{0}$ and  time $t=t'/m$, where $v_{0}= 1$~V, Eq.(\ref{physical}) can be re-written as,
\begin{equation}
\label{system}
\varepsilon \ddot{x} + \gamma \dot{x} + x + a - \frac{g}{\left(x-\dot{x}\right)^2+1}=  \sqrt{2D}n(t), \\
\end{equation}
where $\dot{x}=\frac{dx}{dt}$, $\ddot{x}=\frac{d^2x}{dt^2}$, parameters are $\varepsilon= \frac{CL}{m^2}$, $\gamma = \frac{RC}{m}$, $a = V_{a}/v_{0}$, $g=V_{g}/v^{2}_{0}$, $n(t)$ is normalized Gaussian white noise ($<n(t)>=0$, $<n(t)n(t+\tau)>=\delta(\tau)$), and $D$ is the noise intensity. Numerical modelling of the system (\ref{system}) was carried out by integration using the Heun method \cite{manella2002} with the time step $\Delta t=0.001$. 

One can imagine mechanical realization of the system (\ref{system}). One of the simplest mechanical oscillators is a spring-based pendulum [Fig. \ref{fig2} (b)]. In a linear pendulum case a spring-suspended solid is affected by the force of gravity, $\vec{F}_{1}=m\vec{g}$, and the elastic force, $\vec{F}_{2}$, being proportional to displacement $\vec{x}$ from equilibrium: $\vec{F}_{2}=-k\vec{x}$. Taking into account the influence of air resistance assumed to be proportional to the velocity of motion $\vec{F}_{3}=-\gamma \vec{v}$, one can derive an equation of motion in the vectorial form (Newtown's second law): $m\vec{a}=\vec{F}_{1}+\vec{F}_{2}+\vec{F}_{3}$. Then the scalar form of the equation is: $m\ddot{x}+\gamma \dot{x}+kx=0$. Generally, elastic properties of objects can be more complex (see for example a model of hair bundles with negative stiffness \cite{martin2000,martin2003}). Moreover, the spring can be changed to a complex device [Fig. \ref{fig2} (c)] including elastic medium (the medium $S$ in Fig. \ref{fig2} (c)) and a source of energy (is marked by the source E in Fig. \ref{fig2} (c)). Then one can assume that the elastic force nonlinearly depends both on the displacement $x$ and an instantaneous value of the velocity $\dot{x}$ according to the formula: $F_{2}(x,\dot{x})=-x-a+\frac{g}{(x-\dot{x})^2+1}$. The presence of a noise term $\xi(t)$ can be associated with internal fluctuations of the elastic block or random perturbations of a support beam. Then the equation of motion becomes:
\begin{equation}
\label{system-mechanical}
m \ddot{x} + \gamma \dot{x} + x + a - \frac{g}{\left(x-\dot{x}\right)^2+1}=  \xi(t), \\
\end{equation}
which coincides with  Eq. (\ref{system}).

In the context of the general form (\ref{general-eq}), Eq.~(\ref{system}) describes motion in the potential field $U(x,\dot{x})$ as being:
\begin{equation}
\label{u}
\begin{array}{l}
U(x,\dot{x})= \dfrac{x^{2}}{2\varepsilon} + \dfrac{a}{\varepsilon}x- \dfrac{g}{\varepsilon} \text{arctg}(x-\dot{x})+K,
\end{array}
\end{equation}
in the presence of dissipation defined by the factor $\frac{\gamma}{\varepsilon}>0$. The constant $K$ is neglected, $K=0$. An important detail is that the form of the potential at any time is determined by an instantaneous value of the point mass velocity, $\dot{x}$. Any attempts to group the terms of Eqs. (\ref{physical}) and (\ref{system}) in a different way and to derive another dissipation and potential functions give rise to the appearance of second-order discontinuity in the functions of a potential and dissipation, which contradicts physical nature of the system.

\begin{figure}
\begin{center}
\includegraphics[width=0.8\columnwidth]{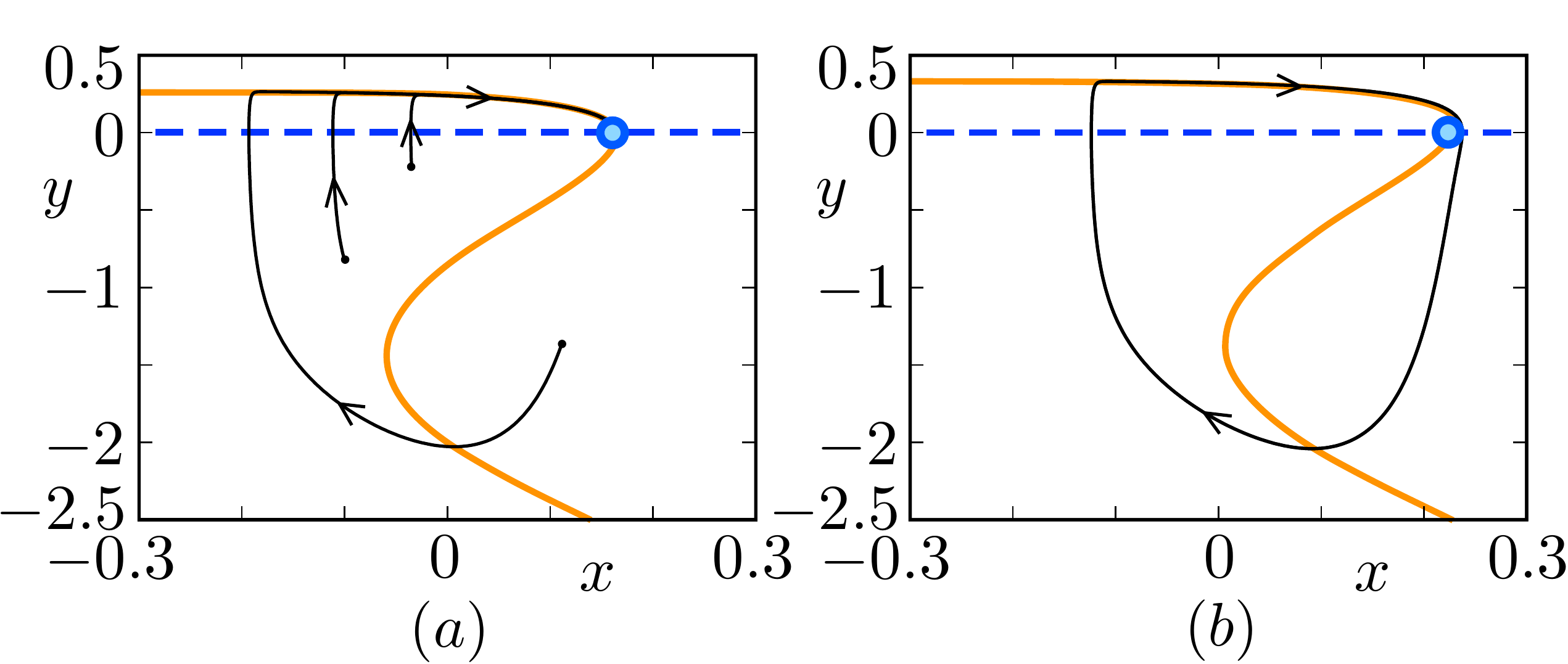}
\end{center}
\caption{Phase space structure of the system (\ref{xy}) in the absence of noise at $a=1.3$ (a) and $a=1.2$ (b). The equilibrium point is shown by the blue circle; the blue dashed line indicates the nullcline $\dot{x}=0$; the orange solid line shows the nullcline ${\dot{y}=0}$. Phase trajectories are shown by black arrowed lines. Other parameters are: $\varepsilon=0.01$, $\gamma=0.5$, $g=1.5$.}
\label{fig3}                                                                                                   
\end{figure}

Let us transform Eq.~(\ref{system}) into a system of two first-order differential equations:
\begin{equation}
\label{xy}
\left\lbrace
\begin{array}{l}
\dot{x}=y,\\
\varepsilon \dot{y}= \dfrac{g}{\left(x-y\right)^2+1}-a-\gamma y- x+\sqrt{2D}n(t).
\end{array}
\right.
\end{equation}
There exists one equilibrium point with the coordinates ($x_{0},y_{0}$) in the phase space of the system~(\ref{xy}) in the absence of noise:
\begin{equation}
\label{equilibrium}
\begin{array}{l}
y_{0}=0,\\
x_{0}=-\frac{a}{3}+\sqrt[3]{-\frac{q}{2}+\sqrt{\frac{q^2}{4}+\frac{p^3}{27}}}\\
\\
+\sqrt[3]{-\frac{q}{2}-\sqrt{\frac{q^2}{4}+\frac{p^3}{27}}},\\
\end{array}
\end{equation}
where $q=\frac{2a^3}{27}+\frac{2}{3}a-g$ and $p=-\frac{a^2}{3}+1$.
The fixed point is stable for the parameters set to $g=1.5$, $\gamma=0.5$, $\varepsilon=0.01$, $a>a_{*}$, where $a_{*}\approx1.277$ [Fig.~\ref{fig3}~(a)]. 
Decreasing of the parameter $a$ involves the supercritical Andronov-Hopf bifurcation at $a=a_{*}$. The fixed point becomes unstable and a stable limit cycle appears in the vicinity of the unstable equilibrium [Fig.~\ref{fig3}~(b)]. The self-oscillations represent the fast-slow dynamics like in the FitzHugh-Nagumo model. It includes slow motions along the nullcline $\dot{y}=0$ and the fast ones, when the phase point falls down from the nullcline.

\section{Noise-induced dynamics}
\label{cr}
\begin{figure}
\begin{center}
\includegraphics[width=0.8\columnwidth]{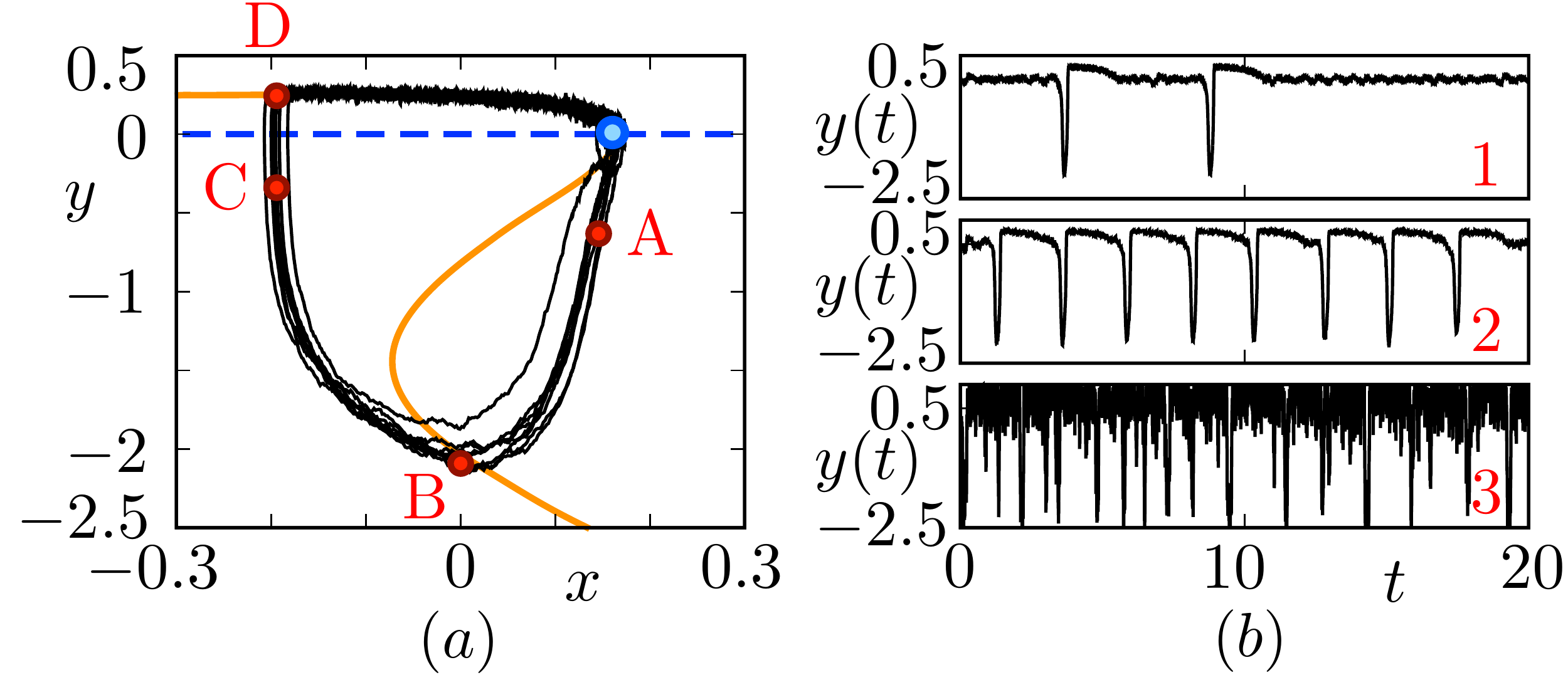}\\
\includegraphics[width=0.8\columnwidth]{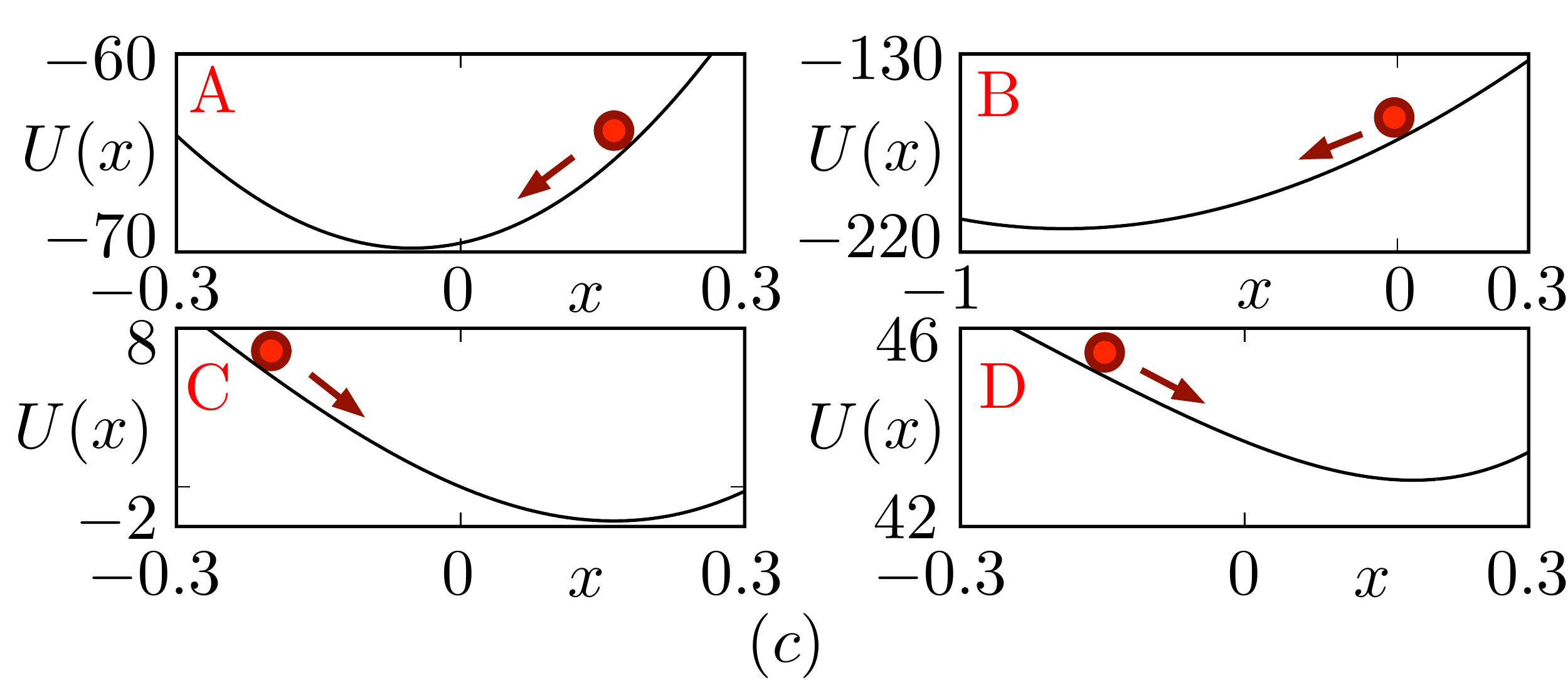}
\end{center}
\caption{System (\ref{xy}). (a) Noise-induced oscillations on phase plane ($x$,$y$) at $D=4\times 10^{-6}$. The equilibrium point is shown by the blue circle; the blue dashed line indicates the nullcline $\dot{x}=0$; the orange solid line shows the nullcline ${\dot{y}=0}$. (b) Time realizations $y(t)$ at $D=4\times 10^{-7}$ (the panel 1), $D=4\times 10^{-6}$ (the panel 2)  and $D= 10^{-3}$ (the panel 3). (c) Potential function $U(x)$ (Eq. (\ref{u})) corresponding to instantaneous velocity in the points A-D on the panel (a). Direction of point mass motion is schematically shown.}
\label{fig4}                                                                                                   
\end{figure}

The system's (\ref{xy}) phase space structure allows to create appropriate conditions for observation of the coherence resonance phenomenon. Next, the parameter $a$ of the system (\ref{xy}) is fixed as being $a = 1.3$, which is close to the Andronov-Hopf bifurcation value. The chosen parameter value corresponds to the existence of a stable fixed point in the phase space of the deterministic system [Fig.~\ref{fig3}~(a)]. In the presence of noise the random force occasionally kicks the phase point out of the vicinity of the stable equilibria. If a stochastic perturbation is sufficient, the phase point traces motion along a loop including a stable branch of the nullcline $\dot{y}=0$ [Fig.~\ref{fig4}~(a)]. The motion along the loop is manifested as spikes in the $y(t)$ time realization [Fig.~\ref{fig4}~(b)]. Small noise [Fig.~\ref{fig4}~(b1)] results in rare spikes with variable interspike intervals. Growth of the noise intensity gives rise to more frequent spiking and increasing of the spiking regularity. There is an optimal value of the noise intensity corresponding to the most regular spiking [Fig.~\ref{fig4}~(b2)] and the almost periodic dynamics. Further increasing of the noise intensity leads to the destruction of regularity [Fig.~\ref{fig4}~(b3)]. 

Description of the system (\ref{xy}) in the context of motion of a particle in a potential field in the control points A-D [Fig. \ref{fig4} (a)] allows to explain the excitability of the system. A stochastic perturbation induces motion of the point mass particle (the red circle in Fig.\ref{fig4} (c)) with non-zero velocity. As a result, position of a local minimum of the potential well changes [Fig. \ref{fig4} (c), inset A] according to the formula (\ref{u}). It can accelerate the particle's motion [Fig. \ref{fig4} (c), inset B]. Then the growth of the velocity decelerates because of a lower slope of the potential function and the presence of positive dissipation. It leads to inverse transformation of the potential function [Fig. \ref{fig4} (c), inset C] and stops the particle. After that the point mass particle comes to its initial state [Fig. \ref{fig4} (c), inset D].

\begin{figure}
\begin{center}
\includegraphics[width=0.8\columnwidth]{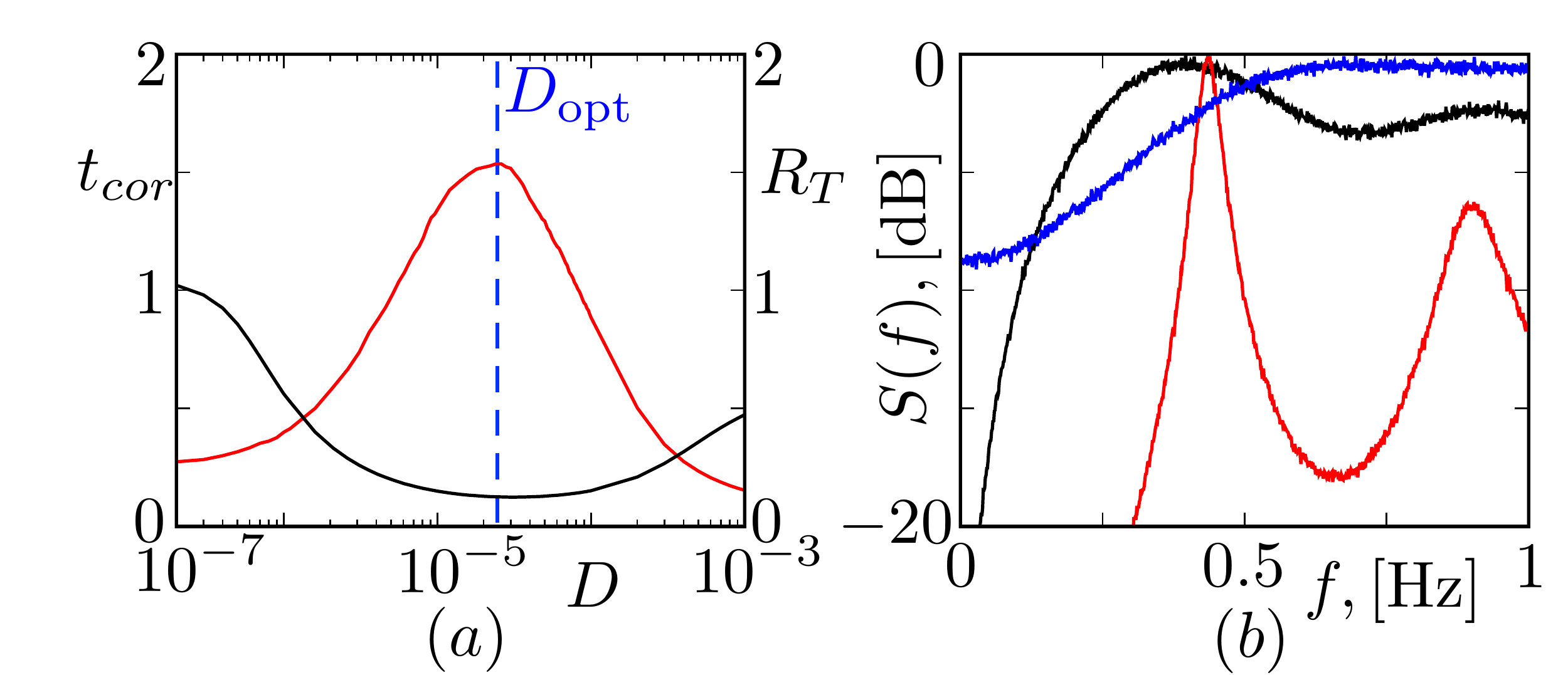}\\
\includegraphics[width=0.8\columnwidth]{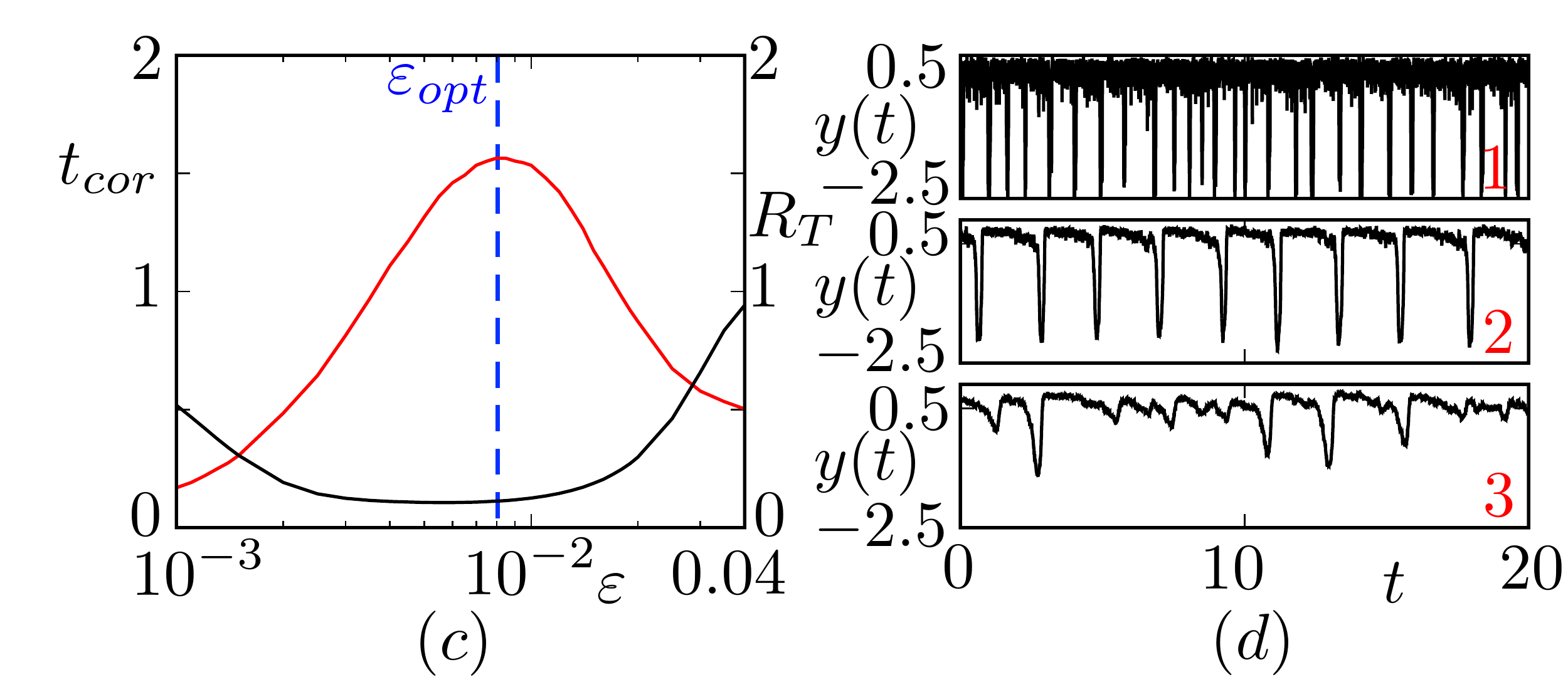}
\end{center}
\caption{System (\ref{xy}). (a) Dependences of correlation time, $t_{cor}$ (the red line), and normalized standard deviation of interspike intervals, $R_{T}$ (the black line), calculated from $y(t)$ time realizations on the noise intensity, $D$. The blue dashed line indicates the optimal noise intensity $D_{\text{opt}}\approx 2.5\times 10^{-5}$ corresponding to most correlated oscillations. (b) Evolution of the normalized power spectra of $y(t)$ time realizations with noise-intensity growth: $D=4\times 10^{-7}$ (the black line); $D=2.5\times 10^{-5}$ (the red line); $D= 10^{-3}$ (the blue line). (c) Dependences of correlation time, $t_{cor}$ (the red line), and normalized standard deviation of interspike intervals, $R_{T}$ (the black line), calculated from $y(t)$ time realizations, on the parameter $\varepsilon$. The blue dashed line indicates the optimal value $\varepsilon_{\text{opt}}\approx 0.008$ corresponding to most correlated oscillations. (d) Time realizations $y(t)$ at $\varepsilon=0.001$ (the panel 1), $\varepsilon=0.008$ (the panel 2) and $\varepsilon=0.03$ (the panel 3). Parameter values are: $\varepsilon=0.01$, $\gamma=0.5$, $g=1.5$, $a=1.3$ (the panels (a) and (b)) and $\gamma=0.5$, $g=1.5$, $a=1.3$,  $D=2.5\times 10^{-5}$ (the panels (c) and (d)).}
\label{fig5}                                                                                                   
\end{figure}

The described above changes of coherence of the noise-induced dynamics are characterized by the non-monotonic dependences of the correlation time and the normalized standard deviation of interspike intervals on the noise intensity \cite{pikovsky1997}. The correlation time is defined as
\begin{equation}
\label{tcor}
t_{cor}=\dfrac{1}{\sigma^2}\int\limits_{0}^{\infty} \left| \Psi(s) \right|ds,
\end{equation}
where $\Psi(s)$ is the autocorrelation function of the signal $y(t)$ and $\sigma^2 =\Psi(0)$ is its variance. The standard deviation of interspike intervals is calculated as 
\begin{equation}
\label{rt}
R_{T}=\dfrac{\sqrt{\left< T_{ISI}^2\right>-\left< T_{ISI}\right>^2}}{\left< T_{ISI}\right>}.
\end{equation}
The dependences $R_{T}(D)$ and $t_{cor}(D)$ obtained in numerical experiments [Fig. \ref{fig5} (a)] indicate the existence of the optimal noise intensity $D_{opt}\approx 2.5 \times 10^{-5}$ corresponding to a local maximum of the function $t_{cor}(D)$ and a local minimum of the function $R_{T}(D)$. Variable character of coherence with the noise intensity growth results in evolution of a power spectrum of the $y(t)$ time realization [Fig.~\ref{fig5}~(b)]. The main spectral peak becomes most prominent for the optimal noise intensity and then fades out with the noise-intensity growth.

A distinctive feature of coherence resonance in the system (\ref{xy}) has been revealed. It is non-monotonic dependence of the correlation time and the standard deviation of interspike intervals on the parameter $\varepsilon$ [Fig. \ref{fig5} (c)] at fixed noise intensity ($D=D_{opt}=2.5 \times 10^{-5}$). There is the optimal value $\varepsilon=\varepsilon_{opt}\approx 8 \times 10^{-4}$ corresponding to maximal coherence [Fig.~\ref{fig5}~(d2)]. Decreasing [Fig.~\ref{fig5}~(d1)] or increasing [Fig.~\ref{fig5}~(d3)] of the parameter $\varepsilon$ gives rise to weakening of the coherence resonance. 
\section{Conclusions}
\label{conclusions}
The system proposed in the current paper is excitable and exhibits the effect of coherence resonance near the Andronov-Hopf bifurcation (type-II excitability). In terms of motion of a particle in a potential field the system~(\ref{system}) describes the motion in a state-dependent potential in the presence of a positive dissipation factor. It corresponds to mutual interaction between the point mass particle and the potential field. The potential determines the dynamics of the particle, but it changes depending on the particle's velocity.

Self-oscillation excitation and the excitable dynamics result from the existence of a nonlinear function, which is responsible for peculiarities of a potential field.  Taking into account the fact that the excitable behaviour is associated with properties of a potential well, one can identify the described coherence resonance phenomenon as motion of a particle in an excitable potential well. This configuration is different from the classical representation involving the presence of a potential with a fixed profile, and the state-dependent dissipation factor, which possesses negative and positive values and is responsible for the existence of self-sustained oscillations and the excitable dynamics. 

The proposed excitable oscillator is a mathematical model of the systems depicted in Fig. \ref{fig2} (a),(c). The phase space structure of the offered model is typical for systems with type-II excitability, which were experimentally and theoretically considered in the past. From this perspective the effect of coherence resonance described in the manuscript is caused by the same reasons as compared to the previous well-known dynamical systems. For this reason the possibility to observe the described behaviour in real physical experiments seems to be evident. Nevertheless, the issue of experimental exploration remains to be interesting for further studies.

\section*{Acknowledgments}
This work was supported by the Russian Ministry of Education and Science (project code 3.8616.2017/8.9).




\end{document}